\documentclass[aps,prx,twocolumn,superscriptaddress,
 amsmath,amssymb,amsfonts,
 floatfix
]{revtex4-2}
\makeatletter
\def\p@subsection{}
\def\p@subsubsection{}
\makeatother

\pdfoutput=1
\usepackage[titletoc,toc,title]{appendix}
\usepackage{amsmath,amssymb}
\usepackage{graphicx}
\usepackage{color}
\usepackage{rotating}
\usepackage{bm}
\usepackage{setspace}
\usepackage{hyperref}
\hypersetup{
    colorlinks=true,
    citecolor=NavyBlue,
    linkcolor=red,
    urlcolor=NavyBlue,
    bookmarksopen=false,
}
\usepackage{float}
\usepackage{soul}
\usepackage[T1]{fontenc}
\usepackage{bbold}
\usepackage{color}
\usepackage{braket}
\usepackage{physics}
\usepackage{caption}
\usepackage{subcaption}
\usepackage[dvipsnames]{xcolor}

\graphicspath{{figures/}}

\renewcommand{\vec}[1]{\bm{#1}}

\newcommand{\vecl}[2]{\vec{#1}^{(#2)}}
\newcommand{\vecll}[1]{\vec{#1}^{(\ell)}}

\newcommand{\tk}{\tilde{\vec{k}}}
\newcommand{\tq}{\tilde{\vec{q}}}
\newcommand{\tp}{\tilde{\vec{p}}}

\renewcommand{\a}{\alpha}
\renewcommand{\b}{\beta}

\newcommand{\G}{\Gamma}

\renewcommand{\l}{\lambda}

\newcommand{\m}{\mu}
\newcommand{\n}{\nu}

\renewcommand{\t}{\tau}

\newcommand{\w}{\omega}

\newcommand{\x}{\xi}
\newcommand{\z}{\zeta}

\begin{document}

\preprint{APS/123-QED}

\title{Microscopic theory for electron-phonon coupling in twisted bilayer graphene}

\author{Ziyan Zhu}
\email{ziyanzhu@stanford.edu}
\email{ziyan.zhu@bc.edu}
\affiliation{Stanford Institute for Materials and Energy Sciences,
SLAC National Accelerator Laboratory, 2575 Sand Hill Road, Menlo Park, CA 94025, USA}
\affiliation{Department of Physics, Boston College,  Chestnut Hill, MA 02467}

\author{Thomas P. Devereaux}
\email{tpd@stanford.edu}
\affiliation{Stanford Institute for Materials and Energy Sciences,
SLAC National Accelerator Laboratory, 2575 Sand Hill Road, Menlo Park, CA 94025, USA}
\affiliation{
Department of Materials Science and Engineering, Stanford University, Stanford, CA 94305, USA}

\affiliation{Geballe Lab for Advanced Materials, Stanford University, Stanford, CA 94305, USA}


\begin{abstract}
The origin of superconductivity in twisted bilayer graphene -- whether phonon-driven or electron-driven -- remains unresolved, in part due to the absence of a quantitative and efficient model for electron-phonon coupling (EPC). 
In this work, we develop a first-principles-based microscopic theory to calculate EPC in twisted bilayer graphene for arbitrary twist angles without requiring a periodic moir\'e supercell.
Our approach combines a momentum-space continuum model for both electronic and phononic structures with a generalized Eliashberg-McMillan theory beyond the adiabatic approximation.
Using this framework, we find that the EPC is strongly enhanced near the magic angle. The superconducting transition temperature induced by low-energy phonons peaks at $1.1^\circ$ around 1~K, and remains finite for a range of angles both below and above the magic angles. 
We predict that superconductivity persists up to $\sim 1.4^\circ$, where superconductivity has been recently observed despite the dispersive electronic bands~\cite{finney2022,gao2024double}.
Beyond a large density of states, we identify a key condition for strong EPC: resonance between the electronic bandwidth and the dominant phonon frequencies. 
We also show that the EPC strength of a specific phonon corresponds to the modification of the moir\'e potential. 
In particular, we identify several $\Gamma$-phonon branches that contribute most significantly to the EPC, which are experimentally detectable via Raman spectroscopy.

\end{abstract}




\maketitle

\section{Introduction}

Twisted bilayer graphene (tBLG) hosts robust superconducting states with critical temperatures up to $3$\, K at the ``magic angle'' of $\sim 1.1^\circ$~\cite{cao2018unconventional,yankowitz2019,cao2020strange,yankowitz2019,lu_superconductors_2019,saito_independent_2020}.
At this angle, hybridization between the two Dirac cones from the individual graphene layers, rotated by a small twist angle, leads to nearly dispersionless electronic bands, indicative of strong electron correlations~\cite{bistritzer2011moire}. 
At the same time, the system has a large number of phonon branches arising from the folded moiré Brillouin zone, many of which can couple strongly to electrons~\cite{angeli2019,koshino2020,liu2022moirephonon,lu2022phonon}.
The presence of flat bands and abundant phonon modes raises a fundamental question: Is superconductivity in tBLG driven primarily by strong electronic correlations, or can phonons play a key role?

To date, the mechanism underlying superconductivity in tBLG remains unresolved, with experiments providing seemingly conflicting evidence. 
Spectroscopic studies have revealed signatures inconsistent with conventional Bardeen-Cooper-Schrieffer (BCS) theory~\cite{oh2021}, while transport measurements have shown $T_\mathrm{c}$ to be insensitive to Coulomb screening, implying a phonon-mediated mechanism~\cite{liu2021}. More recently, experiments have reported superconductivity at larger twist angles near $1.38^\circ$ and $1.45^\circ$, where the bands are no longer flat, and Coulomb screening appears to suppress $T_\mathrm{c}$ in the later experiment~\cite{finney2022,gao2024double}.
On the theoretical side, both correlation-driven and phonon-mediated mechanisms have been proposed~\cite{dodaro2018phases,xu2018topological,guinea2018,guo2018pairing,liu2018chiral,kennes2018strong,wu2018theory,peltonen2018mft,wu2019phonon,lian2019}. However, fundamental questions remain: Which phonon modes couple most strongly to electrons, and how does this coupling depend on the twist angle?

The lack of a comprehensive microscopic theory of electron-phonon coupling (EPC) in tBLG arises from significant computational challenges.
Near the magic angle, tBLG has approximately 10,000 atoms per unit cell and is generally incommensurate. 
This complexity makes calculating the electronic structures and phonon modes through first-principles calculations computationally prohibitive, even for a single twist angle, let alone systematically surveying a wide range of twist angles. 
In addition, with about $30,000$ phonon modes near the magic angle, it is unclear which phonons are important a priori.
Existing studies often assumed the dispersion relation of monolayer graphene phonon modes~\cite{chen2024,wu2019phonon}, used an empirical interatomic potential~\cite {choi2018,angeli2019,choi2021dichotomy}, or applied a low-energy effective model~\cite{koshino2020}. 
These methods either fail to account for all phonons, in particular the low-energy moir\'e phonons that have strong twist-angle dependence~\cite{birkbeck2024measuring}, or require an exact supercell and are computationally expensive at small twist angles. 

To overcome these challenges, we develop a first-principles-based numerical framework for calculating EPC in tBLG. 
Our model does not require any empirical inputs, and is efficient and generalizable to arbitrary twist angles. 
Using a modified Eliashberg-McMillan approach, we compute the EPC strength across a wide range of twist angles and identify the dominant phonon modes. 
We find that EPC peaks near the magic angle and is sufficient to induce superconductivity with a critical temperature of approximately 1~K. Remarkably, phonon-mediated superconductivity due to low-energy phonons persists up to twist angles of $\sim 1.4^\circ$, despite the large single-particle bandwidth ($\sim$100~meV). This persistence arises from an additional requirement for strong EPC in the non-adiabatic regime: a match between the electronic bandwidth and the dominant phonon frequencies.

The rest of the paper is organized as follows. In Sec.~\ref{sec:structure}, we present the electronic and phonon band structure using our momentum-space continuum model. 
In Sec.~\ref{sec:epc}, we use our model to calculate twist-angle-dependent EPC strength and identify the conditions for strong EPC. We discuss the practical consequences of our findings in Sec.~\ref{sec:summary}.

\begin{figure*}[ht!]
    \centering
    \includegraphics[width=0.8\linewidth]{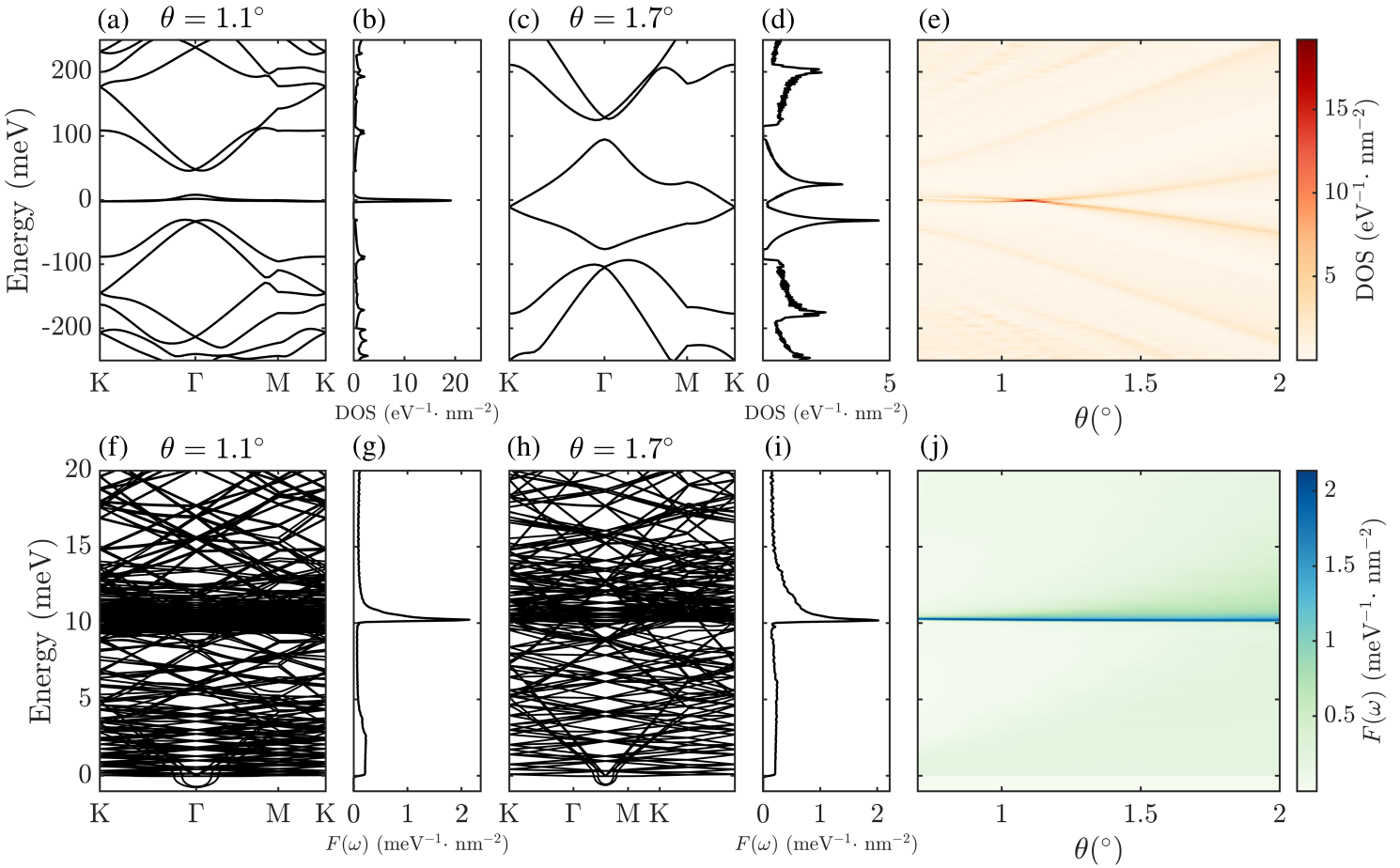}
    \caption{{\bf Electronic and phonon band structures for twisted bilayer graphene. } Electronic band structure and DOS for (a)-(b) $\theta=1.1^\circ$ and (c)-(d) $\theta=1.7^\circ$.(e) Electronic DOS as a function of the twist angle. (f) -(j)  Same as (a) -(e) but for moir\'e phonons.  }
    \label{fig:structure}
\end{figure*}

\section{Momentum-space model for electrons and phonons}\label{sec:structure}

We first obtain the electronic and phonon band structure of tBLG as a function of the twist angle (Fig.~\ref{fig:structure}).
 We treat electrons and phonons on an equal footing via a density functional theory (DFT)-parametrized momentum-space model~\cite{carr2019exact,massatt2023,quan2021phonon,lu2022phonon}, which we review in Appendix~\ref{sec:model}. 
This allows us to easily incorporate the effect of phonons into the electronic structures.
We perform a low-energy expansion and keep only the low-energy degrees of freedom. 
In this way, we retain the computational accuracy of the first-principles calculations without the need for a periodic moir\'e supercell, enabling an efficient twist angle-dependent study. 

In terms of the electronic band structure, changing the twist-angle modifies the interlayer interaction. 
As the twist angle decreases, the two sets of van Hove singularity peaks approach the zero energy (Fig.~\ref{fig:structure}(a)-(e)). 
At a critical twist angle, the magic angle, the density of states (DOS) at the zero energy is enhanced by orders of magnitude and the band becomes nearly dispersionless except for at the $\Gamma$ point.  
Instead of a single magic angle, relaxation leads to a range of twist angles where the electronic density of states has sharp peaks near the zero energy and the precise values depend sensitively on the model parametrization, consistent with previous results~\cite{carr2019exact,bennett2024twist}. 
Using DFT-derived parametrization following \citet{carr2019exact}, the magic angle falls between $1.15^\circ$ and $1.2^\circ$. 
To better match with the experimentally observed magic angle, here and in the rest of the work, we shift the twist angle $\theta$ down by a constant offset $\Delta \theta = 0.1^\circ$, same as in \citet{bennett2024twist}.

The low-energy moir\'e phonon bands are also twist-angle dependent due to a combination of band folding and layer hybridization (Fig.~\ref{fig:structure}(f)-(j)). 
For all twist angles, there is a cluster of dispersionless bands near 10~meV. 
These branches originate from folding the layer breathing mode in Bernal bilayer graphene, which is flat near the $\Gamma$ point. 
At smaller twist angles, more low-energy phonon modes emerge. 
Notably, there are two negative frequency phonon bands near the $\Gamma$-point with $|\omega_\Gamma| < 1\, \mathrm{meV}$. These two modes are shearing modes (Fig.~\ref{fig:negative}).
Given their small negative energies and negligible their density of states (evident in Fig.~\ref{fig:structure}(g) and (i)) and EPC, they are not causes for concern. 
We showed that our model has an excellent agreement with DFT and molecular dynamics in \citet{lu2022phonon}, and the 1.1$^\circ$ low-energy moir\'e phonon band structure and density of states agree with previous works~\cite{angeli2019,liu2022moirephonon}. We include additional computational details in Appendix~\ref{sec:model}.

\section{Electron-phonon coupling in tBLG}~\label{sec:epc}
To quantify the EPC strength, we resort to the Eliashberg-McMillan theory (see Appendix~\ref{sec:mcmillan} for additional details). 
Near the magic angle, the electronic bandwidth is significantly suppressed (on the order of a few meV) while phonons have comparable or even higher frequencies, violating the adiabatic approximation~\cite{sadovskii2019}. 
The EPC constant without the adiabatic approximation is modified as follows, 
\begin{align}
\lambda (E_\mathrm{F}) &= \int \frac{\mathrm{d}\omega}{\omega} \alpha^2 (\omega) F(\omega)  \nonumber \\
&=  \frac{1}{N_\mathrm{F}} \frac{1}{\mathcal{N}_{\tk} \mathcal{N}_{\tq} }  \int \frac{\mathrm{d} \omega}{\omega} \sum_{\tk\tq} \sum_{mn\nu} | g_{mn\nu} (\tk,\tq) |^2 \delta(\omega - \omega_{\tq \nu} )  \nonumber
\\ & \qquad \times \, \delta(\epsilon_{n\tk} - E_\mathrm{F}) \delta(\epsilon_{m\tk  + \tq}- \omega_{\tq \nu}- E_\mathrm{F}),~\label{eqn:lambda1}
\end{align}
where $\alpha^2(\omega) F(\omega)$ is the Eliashberg function, $F(\omega)$ is the phonon density of states, $E_\mathrm{F}$ is the Fermi level, $N_\mathrm{F}$ is the integrated density of states at the Fermi level, $n$ and $m$ are electronic band indices, $\nu$ is the phonon band index, and $\mathcal{N}_{\vec{k}}$ and $\mathcal{N}_{\tq}$ are the number of discretized electron/phonon momenta in the moir\'e Brillouin zone, $\omega$ is the phonon frequency, and $\epsilon_{n\tk}$ is the electronic energy that corresponds to band $n$ at momentum $\tk$. 
The EPC matrix element, $g_{mn\nu} (\tk, \tq)$, is defined as
$
        g_{mn\nu} (\tk, \tq) = \sqrt{\frac{\hbar}{2 M_C \omega_{\tq \nu}} }  \langle \Psi_m (\tk+\tq) | \frac{\mathcal{H}_{\tk} (\vec{r}+\delta\vec{u}_{\tq\nu} (\vec{r})) - \mathcal{H}_{\tk} (\vec{r}) }{|\delta \vec{u}_{\tq\nu} (\vec{r})|} | \Psi_n (\tk) \rangle, \label{eqn:g}
$
where $|\Psi_n\rangle (\tk)$ is the electronic wavefunction at $\tk$, $M_C$ is the mass of carbon atom, $|\delta \vec{u}_{\nu}|$ is the average phonon displacement. We only consider phonon emission since phonon absorption is negligible when we compute at low temperatures. 
We include 6 electronic bands, with 3 above and 3 below the Fermi level and phonon branches up to 20~meV.
In this work, we focus on low-energy phonons because their behavior depends on the twist angle. In contrast, high-energy phonon dispersion remains largely unaffected by changes in the twist angle; the angle dependence of EPC due to high-energy phonons, therefore, mainly arises from variations in electronic bandwidth. 
This choice balances computational feasibility with the goal of isolating the phonon contribution to EPC.
In the rest of the work, we keep the full momentum-dependence of $g_{mn\nu}(\tilde{\vec{k}},\tilde{\vec{q}}).$ 

\subsection{Modification of the moir\'e potential: condition for large EPC}
Computing $\lambda$ using Eq.~\eqref{eqn:lambda1} requires a high-dimensional summation over large number of phonon degrees of freedom, even with a low-energy truncation. 
Instead of directly summing over all phonons, we first attempt to understand which phonons would lead to a large EPC. 
This allows us to exclude phonons that do not contribute to the EPC, eliminating unnecessary calculations of the EPC matrix elements.

\begin{figure*}[ht!]
    \centering
    \includegraphics[width=0.6\linewidth]{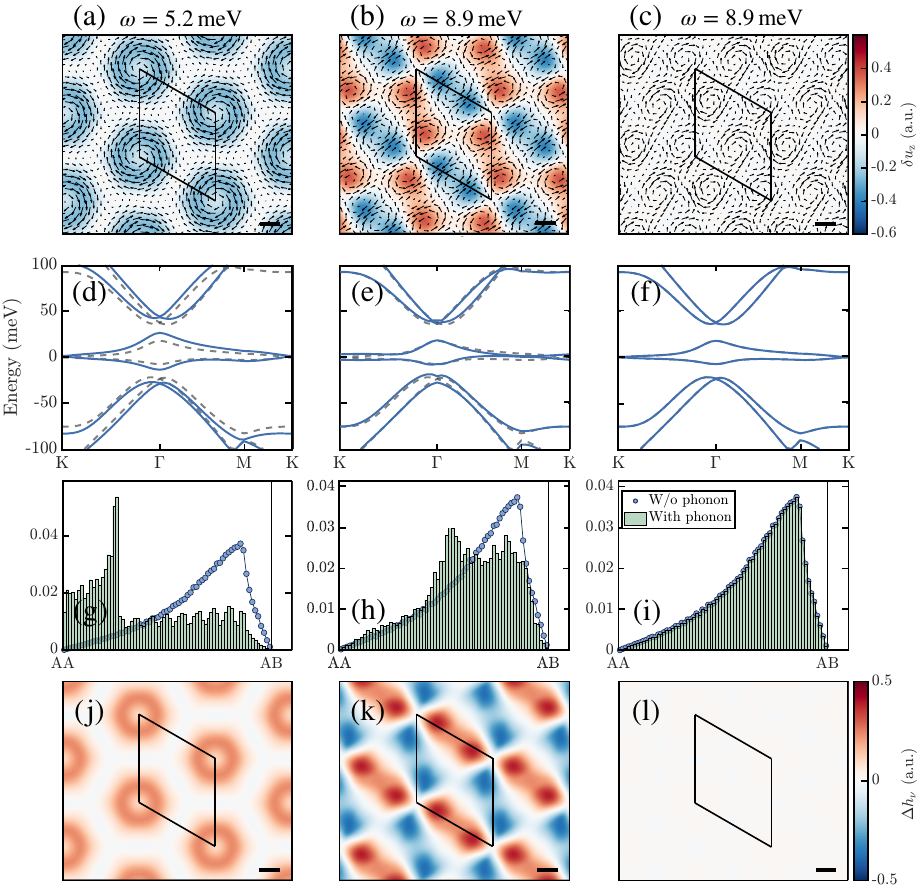}
    \caption{{\bf Comparison between phonons with large and small electron-phonon coupling. }Real space $\Gamma$-point phonon displacement pattern, $\delta \vec{u}_{\Gamma\nu} (\vec{r}) = \delta \vecl{u}{2}_{\Gamma\nu} (\vec{r}) - \delta \vecl{u}{1}_{\Gamma\nu} (\vec{r})$ for (a) $\omega = 5.2$\, meV and (b)-(c)  degenerate $\omega = 8.9\,$meV at $\theta = 1.0^\circ$. Arrows show in-plane displacement vectors, and colors show out-of-plane displacement. The displacements are normalized such that the maximum magnitude is 1. (d)-(f)  Electronic structures with phonons corresponding to (a) -(c), respectively (solid blue lines). Gray dashed lines show the band structure without phonons. (g)-(i)  Redistribution of local stacking order due to phonons. The y-axis shows the probability, and the x-axis shows the distance to AA stacking. Scattered points show the stacking distribution after relaxation and without phonons, which is mostly AB stacking. The histogram shows the stacking distribution with phonons corresponding to (a)-(c), respectively. (j) -(l) Redistribution of the interlayer spacing, $\Delta h_\nu = h(\mathbf{b} + \vec{u}(\vec{b}))  + \delta u_{\mathrm{z},\nu}(\mathbf{b}) - h(\vec{b} + \vec{u} (\vec{b}) + \delta u_{\tq\nu\parallel}(\mathbf{b}))$, that corresponds to the phonons in(a) -(c)  respectively. Scale bars are 30 nm in (a) -(c)  and (j) -(l).  }
    \label{fig:phonons}
\end{figure*}

Intuitively, phonons that substantially modify the moir\'e potential tend to produce large EPC effects.  
We classify changes in the moir\'e potential into two categories: (1) redistribution of stacking configurations due to the in-plane components and (2) change in the interlayer spacing due to the out-of-plane components. 
As examples, Fig.~\ref{fig:phonons} shows three representative low-energy $\Gamma$-point moir\'e phonons (Fig.~\ref{fig:phonons}(a)-(c)) and their impact on the electronic band structure (Fig.~\ref{fig:phonons}(d)-(f)) for $\theta = 1.0^\circ$. 
The phonon shown in Fig.~\ref{fig:phonons}(a) changes the flat band bandwidth and band gap between the flat bands and remote bands. 
The phonon in Fig.~\ref{fig:phonons}(b) opens up a gap between the flat bands. 
The purely in-plane phonon shown in Fig.~\ref{fig:phonons}(c) does not change the electronic structure. 

\subsubsection{In-plane components}

To understand how different phonons affect the system, we compare the distribution of local stacking orders with (green bars) and without (solid lines) phonon displacements, as shown in Fig.~\ref{fig:phonons}(g)–(i). The local stacking order describes the relative alignment of the two graphene layers at each position $\vec{r}$. In the absence of structural relaxation and phonon displacement, this alignment varies smoothly across the moiré supercell.
We quantify the stacking order, $\vec{b}$, by measuring the distance from each local configuration to the AA stacking, where atoms in both layers sit directly on top of each other. Without phonons, the distribution of stacking configurations is skewed toward AB (Bernal) stacking, where one layer is shifted by one-third of the unit cell relative to the other -- an energetically favorable configuration.
In our convention, layer 2 is rotated counterclockwise relative to layer 1. As a result, structural relaxation tends to further rotate the local stacking near AA regions in the same direction, effectively shrinking the high-energy AA domains~\cite{carr2018pressure}.

Phonons can redistribute the equilibrium stacking order in tBLG. 
The phonon mode shown in Fig.~\ref{fig:phonons}(a) has a displacement pattern similar to structural relaxation but rotates in the opposite direction, which results in the expansion of the AA region (Fig.~\ref{fig:phonons}(g)). 
As a result, this phonon influences the band structure in an analogous way to relaxation: it changes the bandwidth at the $\Gamma$ point while preserving the Dirac point degeneracy, consistent with the $\mathcal{C}_6$, 6-fold rotational symmetry, and inversion symmetry that this phonon preserves. 
Normally, the band crossing at the $K$-point is protected by the time-reversal and inversion symmetry of graphene. 
In contrast, the phonon in Fig.~\ref{fig:phonons}(b) does not significantly change the size of the AA spot, but it redistributes some AB stacking areas to lower symmetry stacking configurations (Fig.~\ref{fig:phonons}(h)).
Therefore, the bandwidth at the $\Gamma$-point remains unaffected. 
However, this phonon breaks the $\mathcal{C}_6$ symmetry of tBLG, opening a band gap at the K point, which ceases to be a high-symmetry point. This effect is analogous to the gap induced by strain~\cite{ji2025}.
The phonon in Fig.~\ref{fig:phonons}(c), on the other hand, does not redistribute the local stacking order (Fig.~\ref{fig:phonons}(i)) despite having an in-plane rotation pattern similar to Fig.~\ref{fig:phonons}(b). 
This is because the displacement rotates around fixed stacking orders (primarily AB/BA), effectively translating those domains without changing their type. Therefore, the overall distribution of the local configuration remains unchanged.

\begin{figure*}[ht!]
    \centering
    \includegraphics[width=0.8\textwidth]{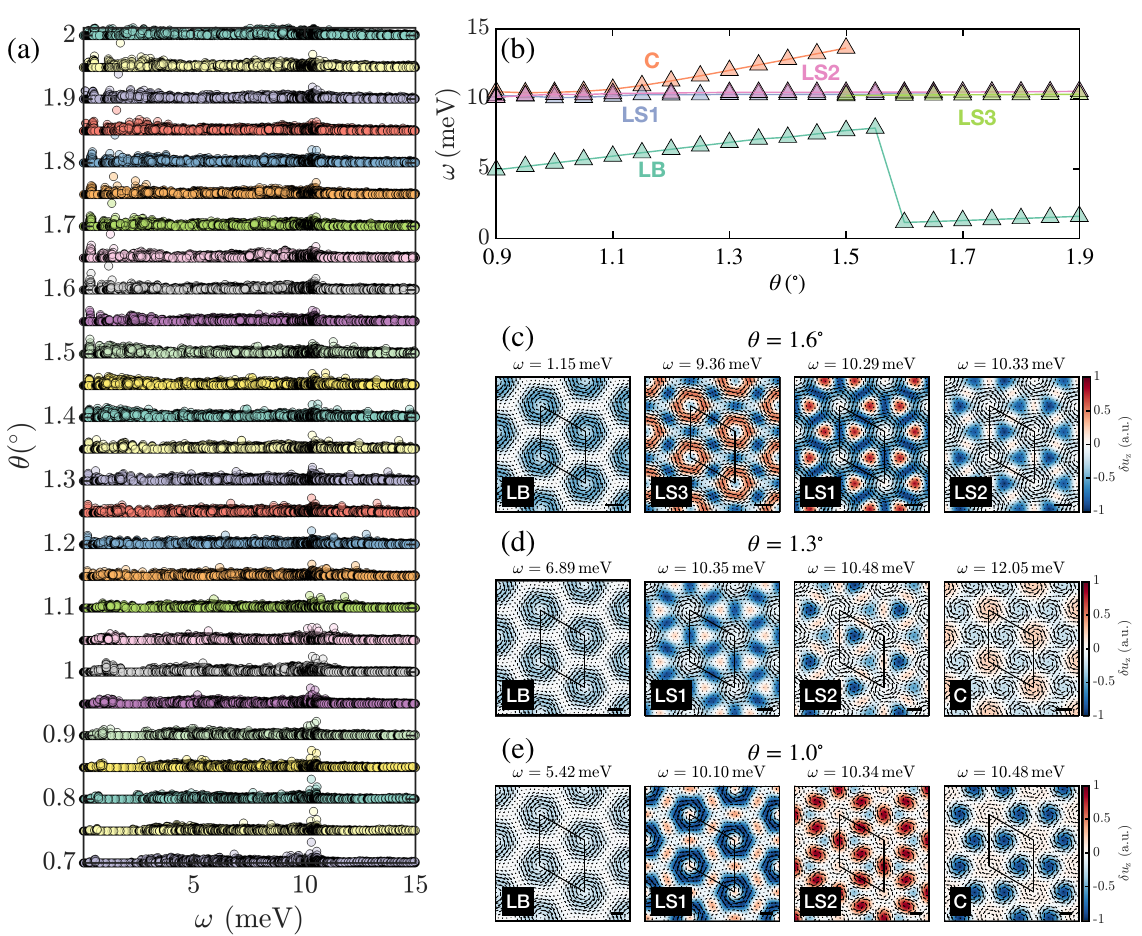}
    \caption{{\bf Identification of phonon branches with large electron-phonon coupling. } (a) Averaged magnitude of the EPC matrix element, $1/{N_{\tk} \mathcal{N}_b^2}\sum_{nm\tk}|g_{mn\nu} (\tk, \tq)|^2$ in arbitrary units, as a function of the phonon frequency for different twist angles. (b) Frequency as a function of the twist angle for $\Gamma$-phonon branches with large EPC identified from(a). Each curve represents a branch labeled by text in the same color: layer breathing (LB), layer shearing (LS), and chiral (C). (c)-(e) Displacement fields, $\delta \vec{u}_{\Gamma\nu} (\vec{r}) = \delta \vecl{u}{2}_{\Gamma\nu} (\vec{r}) - \delta \vecl{u}{1}_{\Gamma\nu} (\vec{r})$, of selective phonons with large EPC at $\theta = $(c) $1.6^\circ$, (d) $1.3^\circ$, (e) $1.0^\circ$. Branch labels are on the lower left corner of each panel. All scale bars are 30 nm. }
    \label{fig:g_vs_theta}
\end{figure*}

\subsubsection{Out-of-plane components}

In addition to the in-plane stacking order, phonons modify the interlayer spacing. 
The equilibrium interlayer spacing associated with a given local stacking configuration, denoted by $h(\vec{b})$, follows a similar functional form to the interlayer misfit energy or generalized stacking fault energy, as discussed in \citet{carr2018relaxation}, with specific coefficients provided in Appendix~\ref{sec:metric}.
In the absence of phonons, the interlayer spacing is largest near the AA spot and smallest near the AB/BA spots. 
When phonons are present, the local stacking vector is perturbed to be $\vec{B} = \vec{b} + \delta \vec{u}_{\tq\nu\parallel}$ where $ \delta \vec{u}_{\tq\nu\parallel}$ denotes the in-plane component of the phonon displacement vector at a phonon momentum $\tq$ and phonon band $\nu$.
If the phonon-modified interlayer spacing, $h(\vec{b}) + \delta u_{\n, z} (\vec{b}) $, deviates significantly from the expected interlayer spacing, $h(\vec{B})$, it can induce a large change in the moir\'e potential. 
For example, the phonon in Fig.~\ref{fig:phonons}(a) exhibits a layer breathing motion, which increases the interlayer spacing near the AA regions. This buckling is again similar to the effect of relaxation, which opens up a band gap between the flat bands and the remote bands~\cite{koshino2017,carr2019exact,fang2019angle} as well as pushes the remote bands to higher energies. 
In contrast, the phonon in Fig.~\ref{fig:phonons}(c) is primarily in-plane, resulting in negligible modification of the interlayer spacing, as shown in Fig.~\ref{fig:phonons}(l).

Combining these two categories, we define a metric that quantifies the total change in the moir\'e potential (Appendix~\ref{sec:metric}). 
We demonstrate that the change of moir\'e potential tracks the magnitude of the EPC matrix element across different phonon branches (Fig.~\ref{fig:metric}). 
In practice, we neglect the phonons with no modification to the moir\'e potential when evaluating the EPC constant (Eq.~\eqref{eqn:lambda1}). This approach eliminates approximately half of the phonons without diagonalizing the phonon-modified Hamiltonian $\mathcal{H} (\vec{r} + \delta \vec{u}_{\tilde{\vec{q}}\nu} (\vec{r}))$, significantly improving computational efficiency.

\subsection{Mode-resolved contributions to EPC}

To investigate the phonons that with the strongest contributions to EPC, we analyze the matrix element $ |g_{mn\nu} (\tk, \tq)|$ averaged over the electronic degrees of freedom $m,n, \vec{k}$, shown in Fig.~\ref{fig:g_vs_theta}(a). 
We observe several phonon branches with strong EPC that exhibit smooth frequency evolution with twist angles. We summarize these phonon branches and their frequency dependence in Fig.~\ref{fig:g_vs_theta}(b). 
All of them are $\Gamma$-phonons that preserve $\mathcal{C}_6$ symmetry. 

There are two branches below 10~meV. 
The higher-frequency branch vanishes near a twist angle of $1.55^\circ$, while another lower-frequency branch emerges around 1.6$^\circ$. 
These two phonon branches have the same real-space displacements (see e.g., the first panel in Fig.~\ref{fig:g_vs_theta}(c)-(e)). 
In particular, their out-of-plane components maintain the same sign across all real-space positions, a characteristic we associate with the layer breathing (LB) mode. 
The reappearance of the LB mode at lower frequency near 1.6$^\circ$ resembles features observed in Raman spectroscopy measurements of twisted bilayer MoS$_2$~\cite{quan2021phonon}.

A pronounced peak in $g$ near 10~meV is universal across all twist angles. 
This behavior is consistent with the twist-angle-independent density of states (Fig.~\ref{fig:structure}(j)) and the strong modulation of the moir\'e potential associated with these modes (Fig.~\ref{fig:metric}). 
These 10~meV phonons are the result of the folding of the untwisted bilayer graphene layer breathing (ZO) mode. 
This 10~meV peak contains three branches, which we collectively refer to as layer shearing (LS) modes. 
The first two modes, LS1 and LS2, have similar in-plane displacement patterns and similar energies. We refer to the slightly lower-frequency branch as LS1 and the slightly higher-frequency branch as LS2. 
Unlike the LB mode, both the in-plane and out-of-plane components of LS1 and LS2 have strong twist angle dependence. The sign of the out-of-plane components varies in real space, resulting in the buckling of the two graphene sheets (Fig.~\ref{fig:g_vs_theta}(c)-(e)). 
The lower-frequency LS1 disappears when $\theta \gtrsim 1.65^\circ$, while LS2 persists across all twist angles. 
At larger twist angle $\theta \gtrsim 1.5^\circ$, an additional LS3 mode appears (see second panel in Fig.~\ref{fig:g_vs_theta}(c)). 
The displacement pattern of LS3 remains relatively insensitive to the twist angle.

An additional distinct branch diverges from the 10~meV peak. Its frequency increases with twist angle and vanishes near $\sim1.5^\circ$. The in-plane displacement of this branch rotates around AA, AB, and BA stacking, all with the same chirality (Fig.~\ref{fig:g_vs_theta}(d)-(e) left panels), distinct from all other branches. We refer to this branch as the chiral (C) mode. On the moir\'e scale, it breaks the inversion symmetry between the AB and BA stackings and can potentially lead to new symmetry-breaking states. 
Chiral phonons have been observed experimentally in monolayer two-dimensional materials such as WSe$_2$~\cite{zhu2018chiral} but not yet in moir\'e materials. 
Previous studies have also reported chiral phonons in tBLG but at the K-point~\cite{liu2022moirephonon}. However, the left and right phonon modes coexist at the K/$\mathrm{K}'$ point, the group velocity vanishes, and therefore all K-point phonons are only local~\cite{chen2021chiral}. The $\Gamma$-point phonon discussed here could support propagation and, due to its strong EPC, offers a promising avenue for experimental detection.

\begin{figure*}[ht!]
    \centering
    \includegraphics[width=\linewidth]{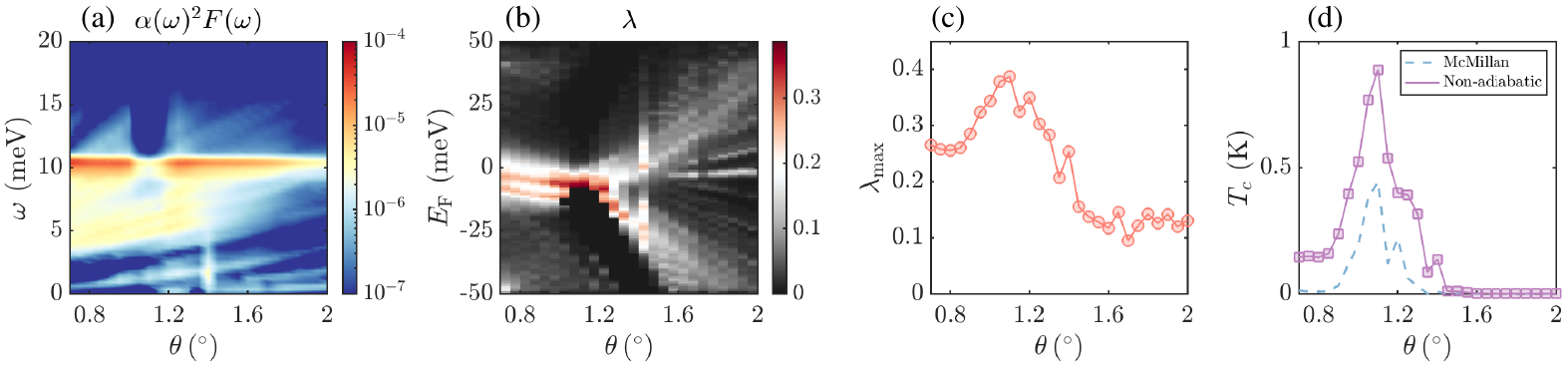}
    \caption{{\bf Twist-angle dependence of electron-phonon coupling and superconducting critical temperature. }(a) $\alpha^2(\omega) F(\omega)$ as a function of phonon frequency and twist angle, averaged over the energy window $E_\mathrm{F} \in [-20, 20]$\, meV. (b) $\lambda$ as a function of the Fermi level, $E_\mathrm{F}$, and the twist angle.
     (c) Maximum $\lambda$ for each twist angle. (d) Estimated critical temperature as a function of the twist angle using McMillan formula (blue dashed line) and McMillan formula (Eq.~\eqref{eqn:tc_anti}) with the finite bandwidth correction (solid purple line).}
    \label{fig:lambda}
\end{figure*}

\subsection{Angle-Dependent EPC and superconductivity}

By combining the contributions of all phonons, we quantify the strength of EPC as a function of the twist angle. 
Figure~\ref{fig:lambda}(a) shows the Eliashberg function, $\alpha^2(\omega)F(\omega)$, averaged over the energy window from $-20$~meV to $20$meV.
We observe two main peaks: one near 10 meV, which remains roughly constant across twist angles, and another at low energies that has increasing frequency with increasing twist angle.

The 10~meV peak originates from phonon branches with both a large DOS and strong EPC.
Despite the weak twist-angle dependence of the matrix element (Fig.~\ref{fig:g_vs_theta}), the 10~meV peak in $\a(\w)^2F(\w)$ dips near $1.1^\circ$, recovers around $1.2^\circ$, and decreases again at larger angles. 
This trend may appear counterintuitive because the electronic DOS has a maximum at $1.1^\circ$ (Fig.~\ref{fig:structure}(e)) and $\l$ is proportional to the DOS. 
However, in the non-adiabatic regime, a high DOS alone does not guarantee enhanced EPC. 
The phonon energy must match available electronic transitions, as enforced by the $\delta(\epsilon_{m\tk+\tq} -\omega_{\tq\nu} - E_\mathrm{F})$ term in Eq.~\eqref{eqn:lambda1}.
Near the magic angles, the electronic bandwidth, $t$, drops below 10~meV, preventing these 10~meV phonons from contributing to the EPC. 
To further confirm this picture, we show the energy difference between the top and bottom flat bands, $t$, as a function of twist angle in Fig.~\ref{fig:bandwidth}.
The twist angles where $\a^2 (\w) F(\w)$ drops align closely with those where $t < 10$\,~meV. 

We integrate $\alpha^2(\omega)F(\omega)$ over frequency to compute the dimensionless EPC constant $\lambda$ for various Fermi energies and twist angles (Fig.~\ref{fig:lambda}(b)). 
Figure~\ref{fig:lambda}(c) presents the maximum $\lambda$ for each twist angle, $\lambda_\mathrm{max}$, which peaks near 0.4. Unlike electronic correlations that decreases rapidly above the magic angle, the EPC remains strong over a broader twist angle range.

We estimate the superconducting critical temperature $T_\mathrm{c}$ using a non-adiabatic formula suitable for narrow bandwidths~\cite{gorkov2016,sadovskii2019,choi2021dichotomy}:
\begin{align}
    T_\mathrm{c} = \prod_i \left( \frac{\omega_i D}{\omega_i + D} \right)^{\l_i/\l} \exp \left(-\frac{1+\tilde{\l}}{\l-\m^\star} \right),\label{eqn:tc_anti}
\end{align}
where $D = t/2$ is the half bandwidth, $\omega_i$ and $\l_i$ are the energy and EPC strength of the i-th phonon mode, $\tilde{\l} = 2\sum_i \l_i D/ (\w_i + D)$ is the mass renormalization constant, and $\mu^\star = \mu/(1+\m\sum_i \ln (1+D/\w_i)^{\l_i/\l})$ is the Coulomb pseudopotential and we take $\m = 0.1$. 
Using the bandwidth-corrected McMillan formula, Fig.~\ref{fig:lambda}(d) shows that $T_\mathrm{c}$ peaks at $1.1^\circ$ with $0.9\,$K, which is similar to the experimental measurements even though we only account for the low-energy phonons~\cite{cao2018unconventional,yankowitz2019,lu_superconductors_2019,saito_independent_2020,oh2021}. 
We find nonzero $T_\mathrm{c}$ across a wide range of twist angles, both above and below the magic angle. 
While the electronic bandwidth remains narrow (on the order of 10~meV) at small twist angles, it increases rapidly beyond the magic angle. 
Coupling to low-energy phonons enables and sustains superconductivity across this range of twist angles, even when the electronic correction is weak. 

To highlight the importance of bandwidth corrections, we also compute $T_\mathrm{c}$ using the conventional McMillan formula, assuming a single phonon mode with $\omega = 10$~meV and $\mu^\star = 0.1$:
\begin{align}
T_\mathrm{c} = \omega \exp \left(- \frac{1+\lambda}{\lambda - \mu^\star(1+\lambda)}\right).
\end{align}
The dashed line in Fig.~\ref{fig:lambda}(d) shows the result. It predicts a lower maximum $T_\mathrm{c}$ and a significantly narrower range of twist angles with superconductivity. 

Here, we briefly comment on our choice of the Coulomb potential, $\mu$. 
The precise value of $\mu$ in TBG is not well established. 
Correlations renormalize the bandwidth, leading to effective energy scales of 40–55~meV near half-filling as measured by STM~\cite{Choi2019,kerelsky2019}, which substantially reduces the ratio $U/t$ compared to single-particle estimates. 
Moreover, while on-site interactions inferred from STM are of order 20~meV~\cite{Xie2019,wong2020}, theoretical studies consistently show that the long-wavelength part of the Coulomb interaction, which is most relevant for pairing, is strongly suppressed by screening near the magic angle~\cite{Goodwin2019,pizarro2019,Vanhala2020}. 
In addition, the effective interaction includes a phonon-mediated attractive contribution, and higher-energy phonons not included in our calculation would further enhance the electron–phonon coupling constant $\lambda$. 
Taken together, these considerations justify our use of $\mu=0.1$ as a reasonable empirical parameter for a qualitative estimate of $T_\mathrm{c}$. 
In addition, $\mu$ is twist-angle dependent, with peaks around the magic angle~\cite{kerelsky2019}. Due to the lack of the exact analytical or numerical form of $\mu$, we keep its value constant. Note that in Eq.~\ref{eqn:tc_anti}, we do account for the twist-angle dependent single-particle bandwidth correction. 

\section{Conclusion and Discussion}\label{sec:summary}

We developed a microscopic EPC model for moir\'e systems that is generalizable for arbitrary twist angles and material combinations. 
Using this framework, we demonstrated that EPC can be strong in tBLG, with large values of $\l$ and enhanced superconducting transition temperatures near the magic angle, which provides direct evidence that phonons can contribute significantly to superconductivity.

We identified low-energy moir\'e phonon modes that strongly couple to electrons, particularly those that modify the moir\'e potential. 
Among them, several branches of $\Gamma$-phonons exhibit large EPC and are potentially observable via Raman spectroscopy. 
Large electron and phonon densities of states are not sufficient for strong EPC.
A critical requirement, especially in the non-adiabatic regime, is the matching between the electronic bandwidth and the phonon frequency. This is true generally when the adiabatic limit is violated. 
As a result of this additional condition, there is a wide range of angles where EPC is enhanced, in contrast to the sharp spike of the electronic density of states. 

While the calculated values of $T_\mathrm{c}$ provide a qualitative trend and indicate where superconductivity may occur, they are not intended to be quantitatively accurate for the following reasons.
First, the Migdal theorem, which neglects vertex corrections in EPC calculations, only holds in two limits: $\omega \gg E_\mathrm{F}$ or $\omega \ll E_\mathrm{F}$~\cite{ikeda1992migdal}. 
Neither limit holds in magic-angle tBLG, so vertex corrections may significantly modify $T_\mathrm{c}$.
Second, we only include the contributions of low-energy phonons because of their pronounced twist-angle dependence~\cite{choi2018,angeli2019,birkbeck2024measuring}. Including higher-energy phonons will further increase the $\lambda$ and $T_\mathrm{c}$ that we estimated. 
Third, the transition temperature is calculated using an s-wave formula, whereas experiments have suggested a nodal nature of superconductivity in TBG~\cite{oh2021,park2025simultaneous}. However, it is well-known that forward-scattering phonons, or phonons that primarily transfer at small momenta, are agnostic to pairing-symmetry, giving an overall boost regardless of nodal structure or not~\cite{johnston2012,lee_interfacial_2014,kozii2019,Rademaker_2016,Bang2019,li_topological_2023}. Due to the absence of an analytical expression for the momentum-dependent interaction $V(\vec{k}, \vec{k}')$, calculating $T_\mathrm{c}$ for a nodal superconductor is left for future work and requires more careful investigation. 

We note that the purpose of this work is to establish a quantitative, twist-angle-dependent and mode-resolved electron-phonon coupling in tBLG. Our weak-coupling $T_\mathrm{c}$ estimate is included only to demonstrate that the computed low-energy EPC can, in principle, support a superconducting instability; it is not intended to account for the full set of unconventional experimental features in tBLG (e.g., V-shaped tunneling spectra~\cite{oh2021}, pseudogaps~\cite{oh2021}, short coherence lengths~\cite{cao2018unconventional}, or intertwined orders such as nematicity~\cite{cao2021nematicity} and Kekul\'e-like correlations~\cite{nuckolls2023quantum}). Capturing these phenomena generally requires going beyond an isotropic Eliashberg-McMillan estimate and treating momentum/orbital structure, strong correlations, and competing orders on equal footing. Our results provide a controlled microscopic input: the twist-angle and mode dependence of EPC matrix elements, which can be incorporated into such future theories, including scenarios where phonons cooperate with electronically driven pairing or intertwined order.

\subsection{Comparison with other EPC models}
We briefly compare our approach with other frameworks for EPC in TBG. 
The most direct route is atomistic modeling~\cite{angeli2019,choi2021dichotomy,liu2022moirephonon}, where, for example, \citet{liu2022moirephonon} combined deep-potential molecular dynamics for phonons with a tight-binding model and the frozen-phonon approximation to obtain phonon-renormalized electronic bands. 
Because the potential is trained on first-principles data, such methods remain close to \textit{ab initio} accuracy. 
However, they are computationally demanding, involve a huge number of degrees of freedom, and are restricted to commensurate twist angles. 
Efficiency can be improved by projecting the atomistic Hamiltonian into a truncated low-energy subspace~\cite{Miao2023truncated}, which preserves atomistic accuracy but still cannot be applied to arbitrary angles. 
A further simplification is provided by continuum models, which expand interlayer hopping or interatomic potentials in small momenta and thus wash out atomic detail~\cite{bistritzer2011moire,koshino2017,carr2019exact,koshino2020,quan2021phonon,lu2022phonon,Vafek2023continuum,xie2023lattice,girotto2023}. 
These models are valid at arbitrary twist angles, and their parameters can be empirical or derived from first-principles calculations. 
For phonons, one can distinguish between (i) continuum elastic models~\cite{koshino2020,xie2023lattice}, which obtain force constants from energy derivatives, and (ii) first-principles--based continuum models~\cite{quan2021phonon,lu2022phonon,girotto2023}, which extract force constants from DFT. 
Our work follows the latter approach. 

In principle, one could also go beyond frozen-phonon calculations using density functional perturbation theory (DFPT), which has been attempted for large-angle TBG~\cite{gao2024}. 
This would require diagonalizing a $3N \times 3N$ dynamical matrix for each phonon and electronic momentum, with $N \sim 11{,}000$ at magic angles, making it computationally prohibitive. 
Wannier interpolation can reduce the cost~\cite{giustino2007}, but still scales poorly with system size and remains impractical for small-angle TBG. 

We emphasize that, despite the variety of available frameworks, our work is the first to quantify the angle dependence of EPC due to phonons and to demonstrate explicitly that phonons can support superconductivity.

\subsection{Experimental consequences and model generalization}

Our findings have several experimental consequences. 
The electronic bandwidth in tBLG can be tuned by electrostatic screening~\cite{guinea2018,liu2021,Cea2022electrostatic,gao2024double}.  
In particular, \citet{gao2024double} shows that Coulomb screening suppresses $T_\mathrm{c}$. 
Since changing the screening also modifies the electronic bandwidth, which in turn modifies the EPC strength through the bandwidth-phonon frequency matching. A suppressed $T_\mathrm{c}$ itself does not rule out the phonon contribution to superconductivity.  
Measuring $T_\mathrm{c}$ as a function of screening and comparing it to phonon characteristics could provide direct experimental insight into the role of phonons in superconductivity. 
Notably, superconductivity has been reported in large-angle tBLG samples ($1.45^\circ$) with $T_\mathrm{c} \sim 0.5$ K, where the bands are no longer flat~\cite{gao2024double}. Our model predicts $T_\mathrm{c} \sim 0.1$ K at $1.4^\circ$ due to low-energy phonons, offering a plausible mechanism for this observation.

The generalizability of our model enables the tuning of superconductivity across a broad range of moiré systems and device configurations. While our analysis focused on low-energy moiré phonons, the same framework can be applied to high-energy monolayer phonons, which are accessible through nano-Raman techniques~\cite{gadelha_localization_2021} as well as angle-resolved photoemission spectroscopy (ARPES)\cite{chen2024}. In monolayer and bilayer graphene, strong EPC arises from the in-plane optical $E_\mathrm{2g}$ modes at $\Gamma$ and the intervalley $A_{1}'$ modes at K, which are responsible for Kohn anomalies and Raman features\cite{piscanec2004,pisana2007,yan2007,ando2006,castroneto2009}. In twisted bilayer graphene, these modes are folded back to the moiré Brillouin zone center, where they can hybridize with moiré phonons and modify both the electronic structure and the superconducting instability. Prior work has suggested that such high-energy phonons can drive valley Jahn–Teller effects~\cite{angeli2019}, stabilize new ground states at filling $\pm 2$ \cite{shi2025moire,wang2025}, and potentially support unconventional superconductivity\cite{wang2024molecular}. Our model can therefore be readily extended to examine the impact of folded monolayer phonons on superconductivity in TBG, an important direction for future studies.

We can extend the model to investigate substrate influences on EPC and their impact on superconductivity. For example, a SrTiO$_3$ substrate enhances the $T_c$ of FeSe thin films through coupling to out-of-plane oxygen phonons~\cite{lee_interfacial_2014}. 
In TBG, the role of an hBN substrate depends sensitively on alignment. When the hBN is misaligned, it primarily suppresses the out-of-plane phonon components while leaving the in-plane modes largely unaffected. Since the phonons that couple most strongly to electrons in our calculation involve both in-plane and out-of-plane motion, the overall correction to EPC is expected to be modest. Intuitively, the substrate increases the rigidity of the graphene sheet against out-of-plane distortions (enhanced Gaussian curvature), an effect more directly relevant for superconductivity. In contrast, when hBN is aligned with TBG, graphene–hBN phonons hybridize strongly and generate additional normal modes, modifying both in-plane and out-of-plane components. This is consistent with ARPES observations of substrate-induced replica bands~\cite{chen2024}.

However, the theoretical understanding of substrate effects on phonons, electronic structure, and $T_c$ remains limited. The flexibility of our model, together with recent experimental advances in graphene–hBN alignment~\cite{hu2023supermoire}, offers a promising framework to address these questions. Due to the lattice mismatch between graphene and hBN, interference between the TBG and graphene/hBN supercells generates a larger moiré-of-moiré pattern that is incommensurate even in the continuum limit~\cite{zhu2020modeling}. To date, neither the relaxed electronic structure nor the phonons have been studied in moiré-of-moiré systems, and extending our model to encapsulated structures warrants further investigation.


Finally, extending our model to transition metal dichalcogenide (TMD)-based moir\'e systems, where superconductivity has been observed with much lower $T_\mathrm{c}$ despite similar electronic DOS enhancement~\cite{xia2024unconventional,guo2024superconductivity}, could reveal material-dependent differences in EPC and eventually clarify the role of phonons in moir\'e superconductivity.

\section*{Acknowledgements}
We thank Stephen Carr, Brian Moritz, Daniel Larson, Mitchell Luskin, Efthimios Kaxiras, Philip Kim, Zhi-Xun Shen, Dunghai Lee, Erez Berg, Leo Li, Elaine Li, Patrick Ledwidth, and Daniel Massatt for their collaborations and helpful discussions. 
This work is supported by the U.S. Department of Energy (DOE), Office of Basic Energy Sciences, Division of Materials Sciences and Engineering.
ZZ is also supported by a Stanford Science Fellowship.  
Computational work was performed on resources of the National Energy Research Scientific Computing Center, supported by the U.S. DOE, Office of Science, under Contract No. DE-AC02-05CH11231. 

\section*{Data Availability Statement}
The dataset used in this work is openly available in \cite{dataset}. 

\appendix

\section{Momentum space model for electrons and phonons}~\label{sec:model}
We frame both moir\'e electron and phonon models on equal footing in momentum space. 
The central idea is to perform the Bloch expansion of an infinite real space model and take a low-energy truncation to ensure a finite basis~\cite{carr2018relaxation,massatt2023,quan2021phonon,lu2022phonon}.
The Hamiltonian can be formally written as a $2\times 2$ block as: 
\begin{equation}
    H_{\mathrm{el/ph}} = \begin{pmatrix}
        H^{11} & H^{12} \\
        H^{21} & H^{22}
    \end{pmatrix}.~\label{eqn:hamiltonian}
\end{equation}
We use superscripts to denote layer index. 
The real space model for the electrons is the tight-binding model with the coupling between different atoms being the hopping parameter:
$
        H^{\ell\ell'}_{\mathrm{el}} = \sum_{\vec{R}^{(\ell)} \vec{R}^{(\ell')} \a \b} c_{\alpha}^\dagger (\vec{R}^{(\ell)}) t_{\alpha \beta} (\vec{R}^{(\ell)} + \vecl{\tau}{\ell}_\a - \vec{R}^{(\ell')} -\vecl{\t}{\ell'}_\b) c_{\beta} (\vec{R}^{(\ell')}) + \mathrm{h.c.},
$        
where $\alpha\beta$ denotes the atomic orbitals, $\t_{\a}$ is the sublattice position, $\vecll{R}$ is an atomic positions on layers $\ell$, and $t_{\a\b}$ is the hopping parameter. 
For phonons, the real space model is the frozen phonon model
$
    H^{\ell\ell'}_{\mathrm{ph}} = \sum_{\vec{R}^{(\ell)} \vec{R}^{(\ell')}} \sum_{\a\b\x\z} b_{\alpha\x}^\dagger (\vec{R}^{(\ell)}) D_{\a\b\x\z} (\vec{R}^{(\ell)} + \vecl{\tau}{\ell}_\a - \vec{R}^{(\ell')} -\vecl{\t}{\ell'}_\b) b_{\beta\z} (\vec{R}^{(\ell')}) + \mathrm{h.c.},
$
where $\x\z$ are the Cartesian degrees of freedom and $D_{\a\b\x\z}$ is the dynamical matrix element. 
Here, we sum over all the atomic positions in both layers and thus do not require an exactly periodic moir\'e supercell. 
Without a twist angle, the Hamiltonian with momenta $\vec{k}$ and $\vec{k} + \vec{G}$, where $\vec{G}$ is a monolayer reciprocal lattice vector, are identical. 
When a twist angle is present and no lattice reconstruction, the diagonal parts are still the monolayer Hamiltonian. 
However, the translational invariance on the monolayer scale is broken, and the two monolayer Hamiltonians at $\vec{k}$ and $\vec{k} + \vec{G}$ can couple through the following interlayer scattering selection rule, with $H^{\ell\ell'} = T^{\ell\ell'}_{\tk}$ for electrons and $H^{\ell\ell'} = D^{\ell\ell'}_{\tq}$ for phonons and $\ell \neq \ell'$:
\begin{widetext}
\begin{align}
      T^{\ell\ell'}_{\tilde{\vec{k}}\a \b} (\vecl{k}{\ell}, \vecl{k}{\ell'}) &= \frac{1}{|\G|} \sum_{\vecl{G}{\ell}\vecl{G}{\ell'}\a\b} e^{i \left( \vecl{G}{\ell}\cdot \vecl{\t}{\ell}_\a - \vecl{G}{\ell'}\cdot \vecl{\t}{\ell'}_\b \right)} \tilde{t}_{\a\b}(\tilde{\vec{k}}+\vecl{G}{\ell}-\vecl{k}{\ell})
       \delta_{\vecl{k}{\ell}-\vecl{k}{\ell'},\vecl{G}{\ell}-\vecl{G}{\ell'}}, \label{eqn:selection_el} \\
    D^{\ell\ell'}_{\tilde{\vec{q}}\a\b\x\z}(\vecl{q}{\ell},\vec{q}^{(\ell')}) &= \frac{1}{|\G|}\sum_{\vec{G}^{(\ell)}\vec{G}^{(\ell')}\a\b\x\z} e^{i \left( \vecl{G}{\ell}\cdot \vecl{\t}{\ell}_\a - \vecl{G}{\ell'}\cdot \vecl{\t}{\ell'}_\b \right)} \tilde{D}_{\a\b\x\z} (\tilde{\vec{q}} + \vec{G}^{(\ell)} - \vecl{q}{\ell}) 
    \delta_{\vecl{q}{\ell}-\vecl{q}{\ell'},\vecl{G}{\ell}-\vecl{G}{\ell'}},\label{eqn:selection_ph}
\end{align}
\end{widetext}
where $|\G|$ is the monolayer unit cell area, $\tk/\tq$ is a momentum in the moir\'e Brillouin zone that the electron/phonon Hamiltonian is centered at, $\vecll{k}/\vecll{q}$ is the electron/phonon momentum-space basis element of layer $\ell$, $\vecll{G}$ is the reciprocal lattice vector of layer $\ell$. 
Equations~\eqref{eqn:selection_el} and \eqref{eqn:selection_ph} impose constraints on the basis elements $\vecll{k}$. 
One choice of the basis elements is that $\vecll{k} = \vecl{G}{\ell'}_{nm} = m \vecl{G}{\ell'}_1 + n \vecl{G}{\ell'}_2$, $\vecl{k}{\ell'} = \vecll{G}_{m'n'} = m' \vecl{G}{\ell}_1 + n' \vecl{G}{\ell}_2$ for $m, n, m', n'\in \mathbb{Z}$ and $\vecll{G}_1$ and $\vecl{G}{\ell}_2$ are the primitive reciprocal lattice vector of the layer $\ell$. 
In other words, the basis elements of layer $\ell$ are the reciprocal lattice vectors of {\it the other layer}. 
This is a natural choice because the atomic position of layer $\ell$ can be represented in relation to the other rotated layer, and the separation between the two layers is the local configuration space. 
While the atomic positions of a twisted system are aperiodic, they are periodic in local configuration. There is a one-to-one mapping between the real and configuration space~\cite{carr2018relaxation,cazeaux2020energy,massatt2023}. 
The mapping between a local configuration $\vec{b}$ and a real space position $\vec{r}$ is $\vec{b}(\vec{r})= (1 - A_1 A_2^{-1}) \vec{r}$, where $A_\ell$ is the matrix with the column vectors being the lattice vectors of layer $\ell$. 

The Fourier transform of the hopping parameter and the dynamical matrix element is defined as $\tilde f(\tp) = \int \mathrm{d}\vec{r}\, e^{i \tp \cdot \vec{r}} f(\vec{r}),$ where $f$ is $t_{\a\b}$ or $D_{\a\b\x\z}$. 
To obtain $\tilde f(\tp)$, we uniformly discretize the configuration space sum over local configurations:
\begin{align}
    \tilde f(\tp) = \frac{1}{N} \sum_{\vec{b}} e^{i \vec{p} \cdot \vec{b}} f(\vec{b}),\label{eqn:rigid}
\end{align}
where $N$ is the number of configurations being summed over and $\vec{p} = 2\pi (1-A_1A_2^{-1})^{-T} \tp$ is a momentum in the monolayer Brillouin zone and note that $\vec{p}\cdot \vec{b} (\vec{r}) = \tp \cdot \vec{r}$. 
The hopping parameter and dynamical matrix element at a configuration $\vec{b}$ is obtained through DFT~\cite{fang2016weakly,fang2018strain,carr2018pressure,lu2022phonon}. 
We use \texttt{phonopy} to analyze the force fields~\cite{phonopy} and \texttt{hiphive} to correct for the anharmonic contributions of the force fields~\cite{hiphive}.

With structural relaxation, atomic positions are modified by the relaxation displacement vector $\vecll{R} \rightarrow \vecll{R} + \vecll{u} (\vecll{R})$.
We use the same relaxation model as in \citet{carr2018relaxation}.. 
The resulting relaxation forms enlarged triangular domains with alternating AB and BA stacking ~\cite{dai2016,koshino2017,carr2018relaxation,gargiulo2018structural,zhang2018structural,yoo2019atomic}. 
As a result, the grid of local configuration is no longer uniform, and thus Eq.~\eqref{eqn:rigid} is modified to be the following,
\begin{equation}
    \tilde f_{\mathrm{relaxed}}(\tk) = \frac{1}{N} \sum_{\vec{b}} e^{i \vec{k} \cdot \vec{b}} f(\vec{b} + \vec{u}(\vec{b})),\label{eqn:relaxed}
\end{equation}
where $\vec{u}(\vec{b}) = \vecl{u}{2}(\vec{b})-\vecl{u}{1}(\vec{b})$. 
In addition, $\vec{u}$ smoothly varies on the moir\'e scale and breaks the translational invariance of the intralayer terms. 
As a result, there is a pseudo gauge field in the electronic Hamiltonian~\cite{koshino2017,fang2019angle,carr2019exact} and a stacking-dependent monolayer term in the phonon Hamiltonian~\cite{lu2022phonon}, which couple the off-diagonal part of the diagonal block in Eq.~\eqref{eqn:hamiltonian}.

\begin{figure}[ht!]
    \centering
    \includegraphics[width=\linewidth]{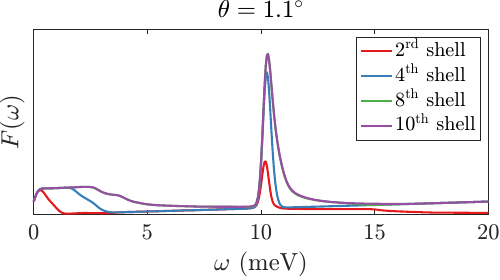}
    \caption{Phonon DOS at $\theta=1.1^\circ$, $F(\w)$, for different momentum space cutoffs shown in different colors. Note that the curve for the $8^\mathrm{th}$-shell (green) overlaps with that of the $10^\mathrm{th}$-shell, suggesting convergence.  }
    \label{fig:dos_scale}
\end{figure}

So far, all equations are exact. For a general twist angle, the basis size is infinite due to incommensurability. 
We now make several approximations to simplify the Hamiltonian. 
We first note that the both the Fourier components of the hopping parameter, $\tilde{t}_{\a\b}$, and dynamical matrix elements, $\tilde{D}_{\a\b\x\z}$, decay rapidly in k-space~\cite{bistritzer2011moire,quan2021phonon}. 
This means that we can keep only the terms in Eqs.~\eqref{eqn:selection_el} and \eqref{eqn:selection_ph} such that $(\vecll{G} - \vecll{k})$ is within some cutoff radius. 
For the electronic structure, if we truncate the sum in Eq.~\eqref{eqn:selection_el} to the first shell of the monolayer reciprocal lattice vectors, ignore the momentum-dependent term in $\tilde t$, $\tilde t_{\a\b} (\tk + \vecll{G} -\vecll{k}) = \tilde t_{\a\b} (\vecll{G}-\vecll{k})$, and neglect the effect of lattice relaxation, we obtain the Bistritzer-MacDonald model~\cite{bistritzer2011moire}. 
However, the twist angle decreases, the strength of relaxation increases, and sharp domain walls form between the neighboring AB/BA stacking below a critical twist angle of $\sim 1^\circ$~\cite{koshino2017,carr2018relaxation,zhang2018structural,yoo2019atomic}.
As a result, a higher-order expansion of $T$ is necessary. 
In our work, we truncate the summation in Eq.~\eqref{eqn:selection_el} to be within $10 |\mathrm{K}|$ where $|\mathrm{K}| = 4\pi/(3 a_0) $ with $a_0 = 2.46$ \AA\ being the monolayer graphene constant. We also keep the $\tk$ dependence of $\tilde t_{\a\b}$ which is known to break the particle-hole symmetry of the flat bands~\cite{carr2019exact,fang2019angle}. 
Phonon bands, on the other hand, are less sensitive to the Fourier expansion of $\tilde D$, and the twist-angle dependence of the phonon bands is primarily a result of band folding with weak hybridization~\cite{angeli2019,liu2022moirephonon}.
Therefore, we approximate $\tilde{D}_{\a\b\x\z} (\tk+\vecll{G}-\vecll{q})= \tilde D (\vecll{G}-\vecll{q})$ and truncate the summation in Eq.~\eqref{eqn:selection_ph} to be within the 8$^{\mathrm{th}}$ shell of the monolayer reciprocal lattice vectors. Figure~\ref{fig:dos_scale} shows the phonon density of states, $F(\w)$, for different cutoff ratios, and confirms that our model has converged at low energies with this cutoff ratio.

\section{Value of the magic angle}
\begin{figure}[ht]
    \centering
    \includegraphics[width=0.9\linewidth]{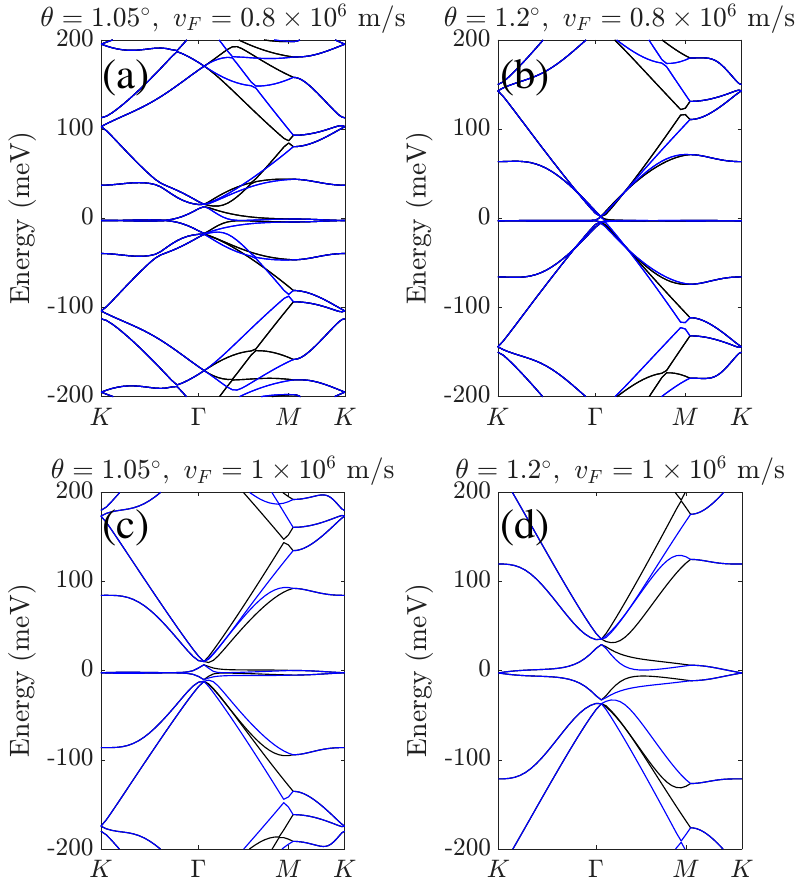}
    \caption{Electronic structures of TBG calculated from the Bistritzer-MacDonald model~\cite{bistritzer2011moire} for two sets of monolayer Fermi velocities (a)-(b) $v_\mathrm{F} = 0.8\times10^6$~m/s and (c)-(d) $v_\mathrm{F} = 1\times10^6$~m/s. Black and blue lines are from $K$ and $K'$ valleys, respectively. }
    \label{fig:vf_compare}
\end{figure}

The precise value of the TBG magic angle depends sensitively on the model parameterization. 
As we mentioned in the main text, using the DFT parameterization~\cite{fang2016weakly,carr2018pressure}, the TBG magic angle occurs at approximately 1.2$^\circ$. 
More explicitly, the Fermi velocity, $v_\mathrm{F}$, obtained from DFT is $0.8\times10^6$~m/s~\cite{fang2016weakly}, whereas $v_\mathrm{F}=1\times10^6$~m/s is extracted from fitting semiclassical calculations to experiments~\cite{castroneto2009}. 
In the Bistritzer-MacDonald model~\cite{bistritzer2011moire} and most subsequent works, $v_\mathrm{F} = 1\times10^6$~m/s is adopted. 
The difference in the $v_\mathrm{F}$ accounts for the shift in the predicted magic angle. 
In Fig.~\ref{fig:vf_compare}, we compare band structures obtained using two different values of Fermi velocities. 
For clarity, in Fig.~\ref{fig:vf_compare} only, we do not include the effect of relaxation, and expand $t (\vec{k})$ to the first shell only, following the original Bistritzer-MacDonald model~\cite{bistritzer2011moire}. 
This is a simpler model than the one we used in the main text. 
With the DFT Fermi velocity, the flattest bands appear at $1.2^\circ$, whereas with the experimental value, they occur at $1.05^\circ$.
In both cases, a magic-angle transition is present, with a narrow bandwidth near $1.1^\circ$. 
In our work, in the spirit of developing an empirical parameter-free framework, we keep the DFT parameters but shift the value of the magic angle to better align with experiments. 

\section{Generalized Eliashberg-McMillan theory for EPC}\label{sec:mcmillan}
We obtain the phonon displacement vector, $\delta \vecll{u}_{\tq\nu}$ at a phonon momentum $\tk$ and a phonon band $\nu$, by diagonalizing the dynamical matrix, and we incorporate $\delta \vec{u}_{\tq\nu}$ into the electronic structure by modifying the atomic position $\vecll{R} \rightarrow \vecll{R} + \vecll{u}(\vecll{R}) + \delta \vecll{u}_{\tq \nu}$. 
When obtaining $\tilde t_{\a\b}$, like how we incorporate the effect of structural relaxation, we modify the summation over uniform $\vec{b}$ in Eq.~\eqref{eqn:rigid} to be over $\vec{b} + \vec{u} (\vec{b}) + \delta \vec{u}_{\tq \nu} (\vec{b})$ with $\delta \vec{u}_\nu (\vec{b}) = \delta \vecl{u}{2}_{\tq \nu} (\vec{b}) -\delta \vecl{u}{1}_{\tq \nu} (\vec{b})$.
Denoting the electronic Hamiltonian at $\tk$ without phonon displacement $\mathcal{H}_{\tk}(\vec{r})$ and with phonon displacement $\mathcal{H}_{\tk} (\vec{r} + \delta \vec{u}_{\tq \nu}(\vec{r}))$, the EPC matrix element, $g_{mn\nu} (\tk, \tq)$, can be calculated as follows,
\begin{widetext}
    \begin{equation}
    g_{mn\nu} (\tk, \tq) = \sqrt{\frac{\hbar}{2 M_C \omega_{\tq \nu}} }  \langle \Psi_m (\tk+\tq) | \frac{\mathcal{H}_{\tk} (\vec{r}+\delta\vec{u}_{\tq\nu} (\vec{r})) - \mathcal{H}_{\tk} (\vec{r}) }{|\delta \vec{u}_{\tq\nu} (\vec{r})|} | \Psi_n (\tk) \rangle, \label{eqn:g}
    \end{equation}
\end{widetext}
where $|\Psi_n\rangle (\tk)$ is the electronic wavefunction at $\tk$, $M_C$ is the mass of carbon atom, $|\delta \vec{u}_{\nu}|$ is the average phonon displacement. 
The diagonal elements of $g_{mn\nu}$ with $m=n$ at $\tq = \Gamma$ are equivalent to calculating the derivative of the electronic energy with respect to the phonon displacement. 


To estimate the electron-phonon interaction, we use the Eliashberg-McMillan approach. 
The standard Eliashberg-McMillan theory of superconductivity is essentially based on the adiabatic approximation in which the phonon frequency is perturbatively small compared to the electronic bandwidth. 
However, for tBLG, especially near the magic angles, the electronic bandwidth is only a few meV (Fig.~\ref{fig:structure}(f)) and many phonons have comparable and even much higher energies (Fig,~\ref{fig:structure}(f)). 
In this case, in addition to the bands at the Fermi level, we need also to include the scattering to remote bands. 
We obtain the general expression for dimensionless EPC constant, $\lambda$, without an adiabatic approximation as follows~\cite{sadovskii2019}:
\begin{align}
\lambda (E_\mathrm{F}) &= \int \frac{\mathrm{d}\omega}{\omega} \alpha^2 (\omega) F(\omega)  \nonumber \\
&=  \frac{1}{N_\mathrm{F}} \frac{1}{\mathcal{N}_{\tk} \mathcal{N}_{\tq} }  \int \frac{\mathrm{d} \omega}{\omega} \sum_{\tk\tq} \sum_{mn\nu} | g_{mn\nu} (\tk,\tq) |^2 \delta(\omega - \omega_{\tq \nu} )  \nonumber
\\ & \qquad \times \, \delta(\epsilon_{n\tk} - E_\mathrm{F}) \delta(\epsilon_{m\tk  + \tq}- \omega_{\tq \nu}- E_\mathrm{F}),~\label{eqn:lambda}
\end{align}
where $\alpha^2(\omega) F(\omega)$ is the Eliashberg function, $E_\mathrm{F}$ is the Fermi level, $N_\mathrm{F}$ is the integrated DOS at the Fermi level, $n$ and $m$ are electronic band indices, $\nu$ is the phonon band index, and $\mathcal{N}_{\vec{k}}$ and $\mathcal{N}_{\tq}$ are the number of discretized electron/phonon momenta in the moir\'e Brillouin zone, $\omega$ is the phonon frequency, and $\epsilon_{n\tk}$ is the electronic energy that corresponds to band $n$ at momentum $\tk$. In this work, we discretize the moir\'e Brillouin zone by $20\times20$ and we include 6 electronic bands and 120 phonon bands. We have verified that changing the discretization and the number of bands included does not change the result.

\section{Negative-frequency phonon modes}~\label{sec:negative}

\begin{figure}
\centering
    \includegraphics[width=\linewidth]{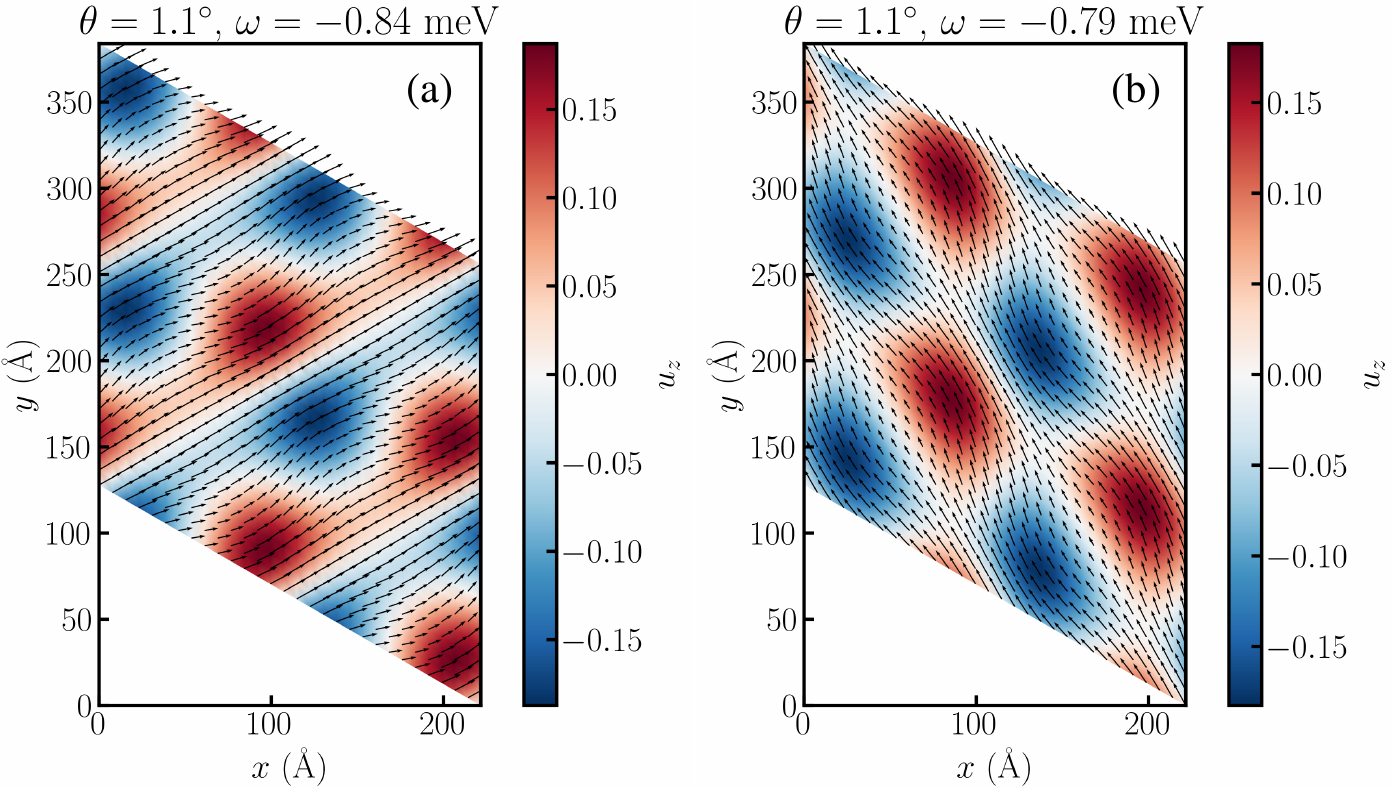}
    \caption{Negative-frequency phonons at $\theta=1.1^\circ$. Arrows are in-plane phonon displacement vectors, and the colors are the strength of the out-of-plane displacement. }
\label{fig:negative}
\end{figure}

In our model, there are two negative phonon branches between $0.7^\circ$ and $2.5^\circ$ near the $\Gamma$-point. 
These modes are numerical artifacts arising from limitations in the underlying DFT calculations
In the stacking dependent calculations, we converge the relaxation to $1\times10^{-6}$\,~eV/\AA \ (additional computational details can be found in ~\citet{lu2022phonon}). 
Specifically, during the stacking-dependent force calculations, the structures are relaxed to a force tolerance of $1\times10^{-6}$~eV/\AA~(see additional computational details in~\citet{lu2022phonon}). When constructing the moiré dynamical matrix, small residual errors from DFT propagate -- particularly because most stacking configurations, aside from AB and BA, are locally unstable.
These negative phonons are shearing modes, with frequencies of less than 1~meV in magnitude. 
Their real-space patterns are shown in Fig.~\ref{fig:negative}: they shift one layer away from each other.
These modes exhibit negligible density of states (see Fig.\ref{fig:structure}(g), (i), (j)) and contribute insignificantly to EPC. We therefore conclude that they do not impact our results.

\section{Quantification of moir\'e potential modification} ~\label{sec:metric}

We quantify the modification of moir\'e potential by combining the effect of both the in-plane and out-of-plane components of a given phonon. 
First, for a given local stacking $\vec{b} = (b_x, b_y)$, define the following vector $\vec{v} = (v , w) \in [0, 2\pi]^2:$
\begin{align}
\begin{pmatrix}
    v\\w
\end{pmatrix} = 2 \pi A^{-T} \begin{pmatrix}
    b_x \\ b_y
\end{pmatrix},
\end{align}
where the column vectors $A$ are the primitive lattice vectors of the monolayer graphene. 
The the interlayer spacing, $h$, as a function of $\vec{b}$, has the following functional form~\cite{carr2018relaxation}:
\begin{align}
    h(v,w) = &c_0 + c_1 [\cos v + \cos w + \cos(v+w)]  \nonumber \\
    &+ c_2[\cos (v+2w) + \cos(v-w) + \cos (2v + w)] \nonumber \\
    &+ c_3 [\cos(2v) + \cos(2w) + \cos(2v + 2w)].
\end{align}
We obtain the coefficients from fitting the interlayer separation from density functional theory calculations at different stacking configurations to be $c_0 = 3.37$ \AA, $c_1 = 0.0209$ \AA, $c_2 = 0.0028$ \AA, $c_3 = 0.0048$ \AA. 

\begin{figure}[ht!]
    \centering
    \includegraphics[width=\linewidth]{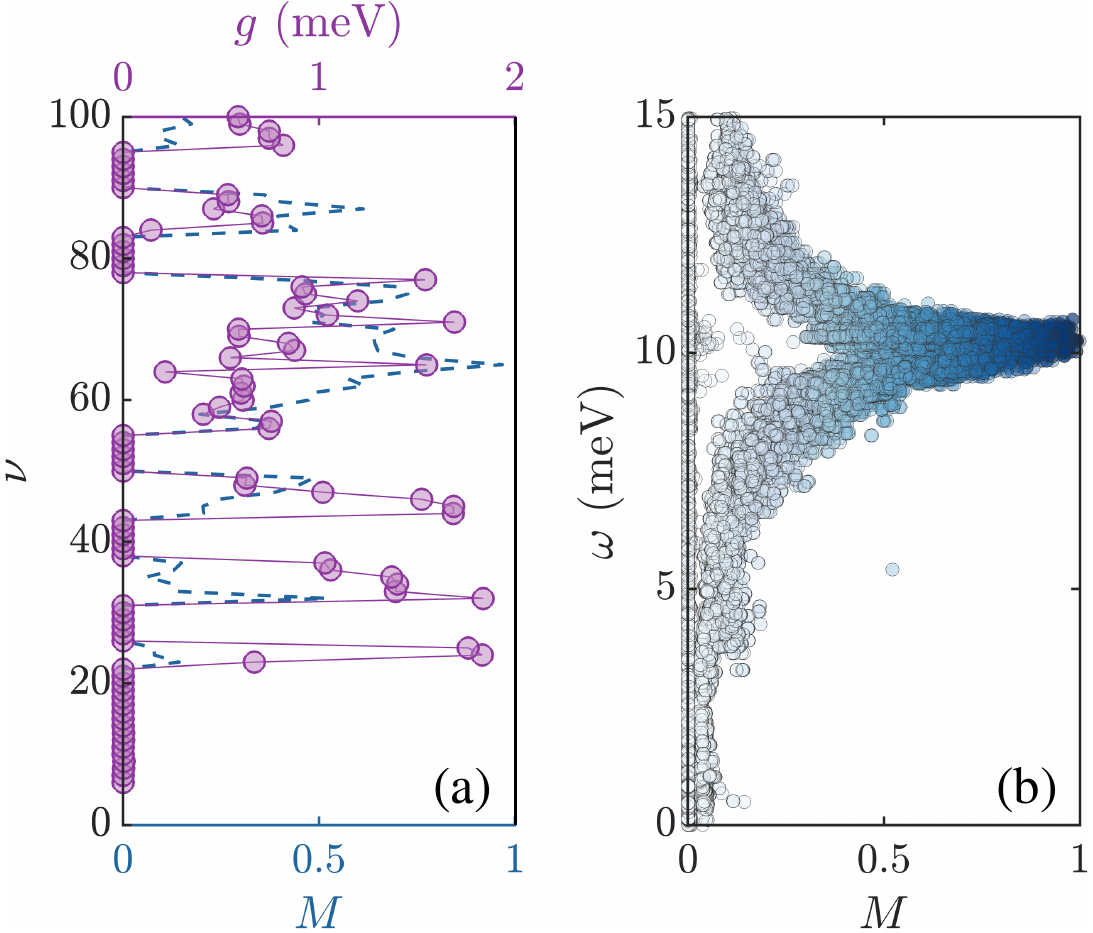}
    \caption{(a) Comparison between the magnitude of EPC matrix element averaged over electronic momenta and bands at $\tq = \G$, $\frac{1}{\mathcal{N}_{\tk}N_\mathrm{b}^2}\sum_{\tk mn}|g_{mn\nu}(\tk,\Gamma)|$ where $\mathcal{N}_{\tk}$ and $\mathcal{N}_\mathrm{b} = 6$ is the number of electronic bands, (purple scattered points, top x-axis) and the EPC (blue dashed line, bottom x-axis) for the 100 lowest phonon bands and with $\theta=1.0^\circ$. (b) EPC metric as a function of the phonon frequency with $\theta=1.1^\circ$ for the 6 lowest energy electronic bands and 100 lowest phonon bands, at all $20\times 20$ electronic and phonon momenta.  }
    \label{fig:metric}
\end{figure}

We define the following metric to quantify the change in the moir\'e potential:
\begin{align}
    M_{\tq\nu} = \frac{1}{2} \sqrt{(D_\mathrm{KL} (P || Q))^2 + \left(\frac{1}{N}\sum_{\vec{b}} \frac{\Delta h_\nu}{\Delta h_{\nu,\mathrm{max}}}\right)^2 },\label{eqn:metric}
\end{align}
where $D_\mathrm{KL}(P || Q) = \sum_{\vec{b}} \log \left( \frac{P(\vec{b})}{Q(\vec{b})} \right)$ is the KL divergence that measures the difference between two distributions $P$ and $Q$, which are the distribution of the stacking order with and without phonons respectively. In addition, $\Delta h_\nu$ is defined as follows: 
\begin{equation}
\Delta h_\nu = h(\mathbf{b} + \vec{u}(\vec{b}))  + \delta u_{\mathrm{z},\nu}(\mathbf{b}) - h(\vec{b} + \vec{u} (\vec{b}) + \delta u_{\tq\nu\parallel}(\mathbf{b})).\label{eqn:deltah}
\end{equation}
The value of $M_{\tq\nu}$ is between 0 and 1. If $M_{\tq\nu}=0$, there is no change in the moir\'e potential, and if $M_{\tq\nu}=1$, there is maximum change to the moir\'e potential.

To confirm that $M_{\tq\nu}$ predicts which phonon has strong EPC, we compare the averaged EPC matrix element at $\G$ (Eq.~\eqref{eqn:g}) over electronic momenta, $\tk$, and electronic bands with $M_{\G\n}$ with $\theta=1.1^\circ$ in Fig.~\ref{fig:metric}(a). 
The metric $M$ has a strong correspondence with $g$, and the magnitude of the EPC matrix element is consistently zero when $M = 0.$

Having established $M$ as a reliable indicator of the EPC matrix element magnitude, we now plot \(M_{\tq\nu}\) for various \(\tq\) and $\nu$, as shown in Fig.\ref{fig:metric}(b). The metric M peaks near 10~meV, where phonons form flat bands. This indicates that the 10~meV phonons strongly modify the moiré potential, which coincides with the large phonon density of states (DOS) in this energy range. These observations confirm that flat phonon bands can substantially alter the moiré potential and give rise to strong EPC.


\begin{figure}[ht!]
    \centering
    \includegraphics[width=\linewidth]{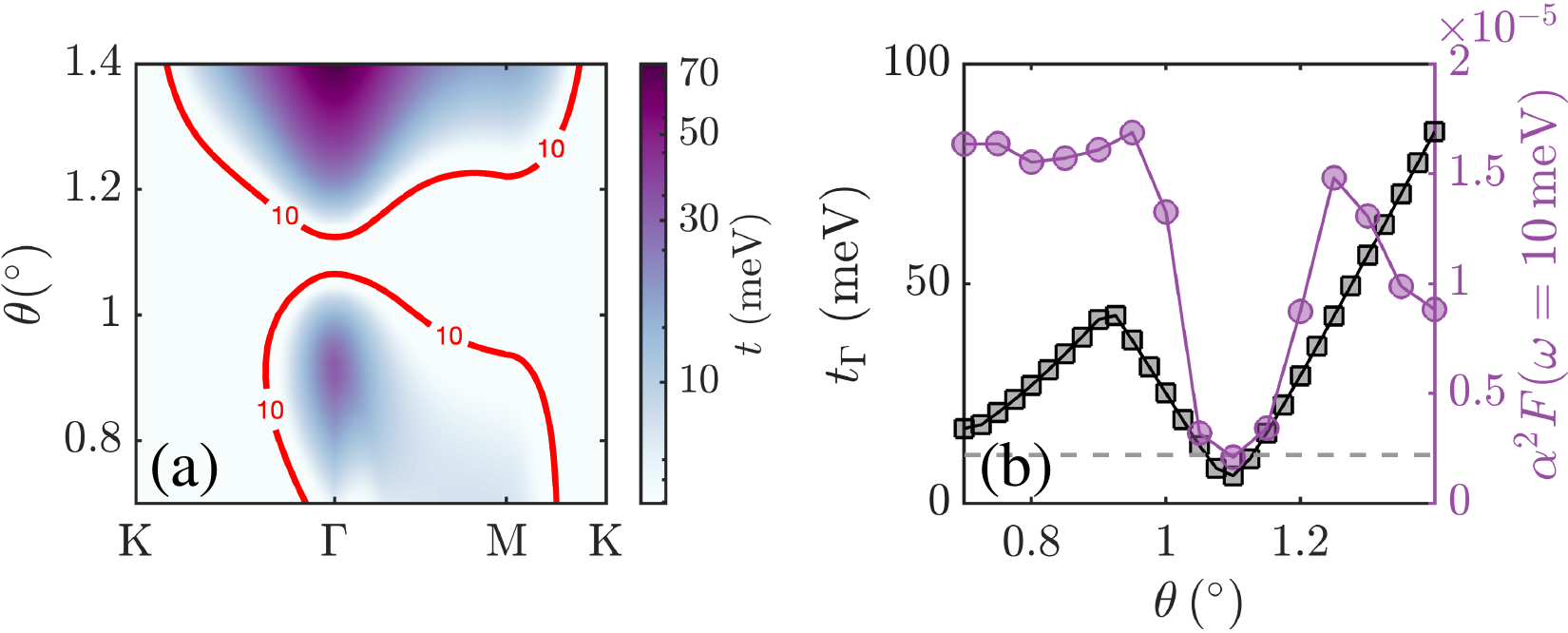}
    \caption{(a) Energy difference between the top and bottom flat bands, $t(\vec{k}) = E_\mathrm{fb,top} (\vec{k}) - E_\mathrm{fb,bot} (\vec{k})$, as a function of the electronic momentum (x-axis) and the twist angle (y-aixs). Red solid lines represent iso-bandwidth lines at 10~meV. (b) Comparison between the bandwidth at the $\Gamma$ point and the Eliashberg function $\alpha^2F(\omega=10~\mathrm{meV})$. Black squares: $\Gamma$-point bandwidth as a function of the twist angle; Grey dashed line: $t_\Gamma = 10$~meV; Purple circles: $\alpha^2F$ at $\omega=10$\,meV.  }
    \label{fig:bandwidth}
\end{figure}

\section{Evolution of bandwidth as a function of the twist angle}
We show the bandwidth as a function of the twist angle in Fig.~\ref{fig:bandwidth}(a). The bandwidth is defined as $t(\vec{k}) = E_\mathrm{fb,top} (\vec{k}) - E_\mathrm{fb,bot} (\vec{k})$, the energy difference between the top and bottom flat bands.
We show iso-$t$ contours at 10~meV in red, which is the phonon frequency with large phonon DOS and strong EPC. 
There is a range of twist angles, between $1.05^\circ$ and $1.15^\circ$, where the bandwidth is always smaller than 10~meV (see the line cut in Fig.~\ref{fig:bandwidth}(b)). 
It is evident from Fig.~\ref{fig:bandwidth}(b) that these twist angles coincide with a dip in the Eliashberg function at 10~meV.
This is because the 10~meV phonons cannot scatter between the flat bands due to the small bandwidth, confirming the resonant condition between the phonon frequency and the electronic energy.

\bibliography{scibib}

\end{document}